\documentclass[english]{iopart}
\begin{document}
\begin{center}
{\Large \bf Connection formulas for the $\lambda$ generalized Ising
  correlation functions}

\author{Barry M. McCoy}

\address{CNYang Institute of Theoretical Physics

 Stony Brook University

Stony Brook, NY, 11790, USA

mccoy@max2.physics.sunysb.edu
}

\end{center}
\begin{abstract}
We derive and prove the connection formulas for the $\lambda$
generalized diagonal Ising model correlation functions.
\end{abstract}

\section{Introduction}

The $\lambda$ generalized Ising model correlations may be defined for
$T<T_c$ by the Fredholm determinant expression \cite{lm}
\begin{eqnarray}
\hspace{-.4in}C^-(N,t;\lambda)
=(1-k^2)^{1/4}{\rm exp}\sum_{n=1}^{\infty}\lambda^{2n}F^{(2n)}_N
\label{fred1}
\end{eqnarray}
\begin{eqnarray}
\hspace{-.4in}F_N^{(2n)}=\frac{(-1)^{n+1}}{n(2\pi)^{2n}}\oint 
\prod_{j=1}^{2n}\frac{dz_j~z_j^{N}}{1-z_jz_{j+1}}\prod_{j=1}^n
Q(z_{2j-1})Q(z_{2j-1}^{-1})P(z_{2j})P(z_{2j}^{-1})
\label{fred2}
\end{eqnarray}
where the contours of integration are 
$|z_j|=1-\epsilon,~z_{2n+1}\equiv z_1$ and 
\begin{equation}
P(z)=1/Q(z)=(1-kz)^{1/2}
\end{equation}
The parameter $k$ satisfies $0\leq k \leq 1$ and we will use
\begin{equation}
t=k^2
\end{equation}

It should be noted that an equivalent form of
(\ref{fred1})-(\ref{fred2}) was given in the 1976 paper of
\cite{wmtb}. 
The integrals of the two expressions are presumably seen to be equal by adding
appropriate total derivatives to the integrands but this has never been
explicitly demonstrated.

When $\lambda=1$ the Fredholm determinant (\ref{fred1}) reduces 
to the diagonal correlation function of
the Ising model given by the $N\times N$ Toeplitz determinant \cite{book}
 \begin{eqnarray}
C(N,t;1)=
\begin{array}{|llll|}
{a}_0&{a}_{-1}&\cdots&{a}_{-N+1}\\
{a}_1&{ a}_0&\cdots&{a}_{-N+2}\\
\vdots&\vdots&&\vdots\\
{a}_{N-1}&{a}_{N-2}&\cdots&{a}_0
\end{array}
\label{detdn} 
\end{eqnarray}
with
\begin{eqnarray}
a_n={1\over 2\pi}\int_{0}^{2\pi}d\theta
e^{-in\theta}\left[\frac{1-ke^{-i\theta}}
{1-ke^{i\theta}}\right]^{1/2}
\label{dn3}
\end{eqnarray}
and
\begin{eqnarray}
k=(\sinh 2E_v/kT \sinh 2 E_h/kT)^{-1}
\nonumber
\end{eqnarray}

The expression (\ref{fred1})-(\ref{fred2}) is obtained in \cite{lm} 
for $\lambda=1$ by extending to
all orders the proceedure used by Wu \cite{wu} to compute  the leading order
expansion of (\ref{detdn}) for $k$ fixed and $N\rightarrow \infty$.

The generalized correlation (\ref{fred1}) for $0<t<1$ satisfies the 
sigma form of the
Painlev{\'e} VI equation first derived by Miwa and Jimbo \cite{jm} for the
diagonal Ising correlation
\begin{equation}
\hspace{-.4in}\left(t(t-1)h''\right)^2
+4h'\left( (t-1)h'-h-1/4\right)\left(th'-h\right)
=N^2\left((t-1)h'-h\right)^2
\label{jm}
\end{equation}
where
\begin{equation}
\hspace{-.4in}h(t)=t(t-1)\frac{d}{dt}\ln C^{-}-\frac{t}{4}~~{\rm with}~~  
t=(\sinh 2E_v/kT\sinh 2E_h/kT)^{-2}
\label{hmcm}
\end{equation}
It is readily seen from the definition (\ref{fred1}) that these
generalized correlations are analytic at $t=0$ and that for
$t\rightarrow 0$
\begin{equation}
C^-(N,t;\lambda)=(1-t)^{1/4}\{1+\lambda^2\frac{(1/2)_N(3/2)_N}
{4[(N+1)!]^2}t^{N+1}(1+O(t))\}
\label{bcm}
\end{equation}
where by noting that
\begin{equation}
t(t-1)\frac{d}{dt}\ln (1-t)^{1/4}-\frac{t}{4}=0
\end{equation}
we have for $t\rightarrow 0$ the one parameter boundary condition for (\ref{jm})
\begin{equation}
h(t)\rightarrow -\lambda^2t^{N+1}\frac{(1/2)_N(3/2)_N}{4N!(N+1)!}
\label{hmbc}
\end{equation}

The question of interest is to determine the behaviour of this one
parameter family at $t=1$. A local analysis of the nonlinear equation
(\ref{jm}) gives the result that as $t\rightarrow 1$ that
\begin{eqnarray}
&&h(t)=-\frac{1-\sigma^2}{4}+\frac{1}{8}(1-\sigma^2)(1-t)\nonumber\\
&&\hspace{.5in}+\frac{s}{16\sigma}(1+\sigma)(2N+\sigma)(1-t)^{1-\sigma}\nonumber\\
&&\hspace{.5in}-\frac{s^{-1}}{16\sigma}(1-\sigma)(2N-\sigma)(1-t)^{1+\sigma}
\label{t1finalh}
\end{eqnarray}
and thus
to order
$(1-t)^{1+\sigma}$
\begin{eqnarray}
&&C^-(N,t;\sigma,{\hat s})=K(N;\sigma)(1-t)^{\sigma^2/4}[1-\frac{1-\sigma^2}{8}(1-t)\nonumber\\
&&\hspace{.8in}+\frac{{\hat s}(N;\sigma)}{16\sigma}
(2N+\sigma)(1-t)^{1-\sigma}\nonumber\\
&&\hspace{.8in}-\frac{{\hat s}(N;\sigma)^{-1}}{16\sigma}
(2N-\sigma)(1-t)^{1+\sigma}+O((1-t)^{2-\sigma})]
\label{cmat1final}
\end{eqnarray}
where $\sigma(\lambda)$ and ${\hat s}(N,\sigma)$ are two integration 
constants for the second order equation (\ref{jm}) and $K(N,\sigma)$ is a
normalizing constant which can  be determined from the original
definition (\ref{fred1}). The computation of
$\sigma(\lambda),{\hat s}(N,\sigma),K(N,\sigma)$
is the purpose of this note. 

\clearpage

The results are as follows
\begin{eqnarray}
&&\hspace{-.6in}\sigma=(2/\pi)\arccos \lambda
\label{res1}\\
&&\hspace{-.6in}
{\hat s}(N,\sigma)=16^{\sigma}\prod_{n=1}^N\frac{1-\sigma/2n}{1+\sigma/2n}
\label{res2}\\
&&\hspace{-.6in}
K(N;\sigma)=2^{-\sigma^2}\left(\frac{\sigma}{\sin\pi\sigma/2}\right)^N
\prod_{m=1}^{N-1}\left(1-\frac{1}{4m^2}\right)^{m-N}
\prod_{m=1}^{N-1}\left(1-\frac{\sigma^2}{4m^2}\right)^{N-m}
\label{res3}
\end{eqnarray}

In section 2 we briefly discuss the history of this connection problem
and in section 3 we present special cases which confirm the results
(\ref{res1})-(\ref{res3}). A proof of (\ref{res2}) and (\ref{res3}) is
given in section 4 by use of the Toda-like relation of Mangazeev and
Guttmann \cite{mangazeev}.

\section{History}

In the scaling limit 
\begin{equation}
t\rightarrow 1,~~~N\rightarrow \infty,~~~{\rm with}~~N(1-t)=r~~{\rm
  fixed}
\end{equation}
the scaling function
\begin{equation}
G_{-}(r;\lambda)=\lim_{\rm scaling}(1-t)^{-1/4}C^-(N,t;\lambda)
\end{equation}
was shown in 1976 in \cite{wmtb} for $\lambda=1$ to be expressed in terms of a Painlev{\'e}
III function and the connection formulas for $\sigma(\lambda)$ and
${\hat s}(N,\sigma)$ for the $\lambda$ generalized
scaling function were computed the next year in \cite{mtw} by a direct
expansion of the integrals in the scaled version of
(\ref{fred1})-(\ref{fred2}). The normalization constant was 
computed in 1991 by Tracy in
\cite{tracy}.

For the generic case of the Painlev{\'e} VI function the connection
formulas for $\sigma$ and ${\hat s}$ were computed in 1982 by Jimbo 
in \cite{jimbo} by means of
deformation theory and  the normalizing constant $K$ was computed in
2018 by Its, Lisovyy and Prokhorov \cite{its}.

In the generic
case there is nonanalytic
behavior at all three points $t=0,1,\infty$. However, the Painlev{\'e}
VI for the Ising model is not generic and therefore the results
of \cite{jimbo}, while still relevant at $t=1$ where the correlation
function is singular, do not hold at
$t=0,\infty$ where the correlation function is analytic. 
Nongeneric cases have been studied by Guzzetti
\cite{guzzetti}) but  the case relevant for
the generalized Ising correlations seems   not to have  been investigated.  

\section{Special Cases}

We here present several special cases of computations which confirm the
results (\ref{res1})-(\ref{res3}).

\subsection{$C^-(N,t;\lambda)$ for $N=0,1$}

In \cite{ongp} a prescription is given to express $C^-(N,t;\lambda)$
in terms of the theta functions
\begin{eqnarray}
&&\theta_2(u;q)=2\sum_{n=0}^{\infty}q^{(n+1/2)^2}\cos[(2n+1)u]\label{theta2def}\\
&&\theta_3(u;q)=1+2\sum_{n=1}^{\infty}q^{n^2}\cos2nu\label{theta3def}\\
&&\theta_4(u;q)=1+2\sum_{n=1}^{\infty}(-1)^nq^{n^2}\cos2nu\label{theta4def}
\label{thetadef}
\end{eqnarray}
where $q$ is related to the variable $t$ by
\begin{equation}
q=e^{-\pi K'(k)/K(k)}
\label{qdef}
\end{equation}
with $k^2=t$ and $K(k)$ and $K'(k)$ are the complete elliptic integrals
of the first kind. 
In \cite{boukraa} and \cite{mccoyetal} this prescription was used to
obtain explicit expressions for $N=0,1$
\begin{eqnarray}
&&C^-(0,t;\lambda)=\frac{\theta_3(u;q)}{\theta_3(0,q)}\label{c0}\\
&&C^-(1,t;\lambda)=\frac{-\theta'_2(u;q)}{\sin(u)\theta_2(0,q)\theta_3(0;q)^2}
\label{c1}
\end{eqnarray}
where prime indicates the derivative with respect to $u$ and
\begin{equation}
\lambda=\cos u
\end{equation}

The expressions  (\ref{c0}) and (\ref{c1}) are expanded at $t=0$ 
by the direct use of (\ref{theta2def}) and (\ref{theta3def}) to obtain
the form (\ref{bcm}). The expansion at $t=1$ is
obtained from (\ref{c0}) and (\ref{c1}) by use of the identities
(on page 370 of \cite{bateman2} with $v\rightarrow u/\pi)$
\begin{eqnarray}
&&\hspace{-.8in}
\theta_3(\frac{uK(k)}{i K'(k)};e^{-\pi K(k)/K'(k)})e^{ -u^2K(k)/\pi K'(k)}
=\left(\frac{K'(k)}{K(k)}\right)^{1/2} 
\theta_3(u;e^{-\pi K'(k)/K(k)})\\
&&\hspace{-.8in}
\theta_4(\frac{uK(k)}{i K'(k)};e^{-\pi K(k)/K'(k)})e^{ -u^2K(k)/\pi K'(k)}
=\left(\frac{K'(k)}{K(k)}\right)^{1/2} 
\theta_2(u;e^{-\pi K'(k)/K(k)})
\end{eqnarray}
The results (\ref{res1})-(\ref{res3}) are obtained by comparing these
explicit expansions at $t=1$ with the general form (\ref{cmat1final}).

\subsection{Algebraic cases for $\lambda=\cos(m\pi/n)$ where
  $\sigma=2m/n$
}

When 
\begin{equation}
\lambda=\cos m\pi/n
\end{equation}
the function $C^-(N,t;\lambda)$ is an algebraic function and in
\cite{boukraa}
the explicit results are given for $N=0,1,2$ that for
$\lambda=\cos(\pi/4)$ with $\sigma=1/2$
\begin{eqnarray}
&&\hspace{-.6in}C^-(0,t;\cos(\pi/4))=2^{-1/4}(1-t)^{1/16}[1+(1-t)^{1/2}]^{1/4}\\
&&\hspace{-.6in}C^-(1,t;\cos(\pi/4))=2^{-3/4}(1-t)^{1/16}[1+(1-t)^{1/2}]^{3/4}\\
&&\hspace{-.6in}C^-(2,t;\cos(\pi/4))=2^{-5/4}(1-t)^{1/16}[1+(1-t)^{1/2}]^{5/4}
[5-(1-t)^{1/2}]/4
\end{eqnarray}

When $t=0$ these expressions are expanded as
\begin{eqnarray}
&&C^-(0,t;\cos(\pi/4))=(1-t)^{1/4}(1+\frac{1}{8}t+\frac{5}{64}t^2
+\frac{15}{256}t^3+O(t^4))\\
&&C^-(1,t;\cos(\pi/4))=(1-t)^{1/4}(1+\frac{3}{128}t^2+\frac{3}{128}t^3+O(t^4))\\
&&C^-(2,t;\cos(\pi/4))=(1-t)^{1/4}(1+\frac{5}{516}t^3+O(t^4))
\end{eqnarray}
which agree with the expansion at $t=0$ of (\ref{bcm}) with $\lambda
=\cos(\pi/4)=2^{-1/2}$.

When $t\rightarrow 1$ these expressions are expanded as
\begin{eqnarray}
&&C^-(0,t;\cos(\pi/4))=2^{-1/4}(1-t)^{1/16}[1+\frac{1}{4}(1-t)^{1/2}\nonumber\\
&&\hspace{1in}-\frac{3}{32}(1-t)+\frac{7}{128}(1-t)^{3/2}+O(t^2)]\label{cm1o40}\\
&&C^-(1,t;\cos(\pi/4))=2^{-3/4}(1-t)^{1/16}[1+\frac{3}{4}(1-t)^{1/2}\nonumber\\
&&\hspace{1in}-\frac{3}{32}(1-t)+\frac{5}{128}(1-t)^{3/2}+O(t^2)]\label{cm1o41}\\
&&C^-(2,t;\cos(\pi/4))=2^{-5/4}\frac{5}{4}(1-t)^{1/16}
[1+\frac{21}{20}(1-t)^{1/2}\nonumber\\
&&\hspace{1in}-\frac{3}{32}(1-t)-\frac{9}{128}(1-t)^{3/2}+O(t^2)]\label{cm1o42}
\end{eqnarray}

The only other case where an explicit result is given in
\cite{boukraa} is for
\begin{equation}
\lambda=\cos(\pi/3)=\frac{1}{2} ~~{\rm with}~~\sigma=2/3
\end{equation}
where for $N=0$ the function $C^-(0,t;\cos(\pi/3)$ satisfies 
\begin{equation}
16C^{12}-16C^9+8t(1-t)C^3+t(1-t)=0
\label{m1n3eqn}
\end{equation}
The expansion near $t=0$ of the solution of  (\ref{m1n3eqn})  
which does not vanish at $t=0$ is  
\begin{equation}
C^-(0,t;\cos(\pi/3))=(1-t)^{1/4}[1+\frac{1}{16}t+O(t^2)]
\end{equation}
and for $t\rightarrow 1$
\begin{eqnarray}
&&C^-(0,t;\cos(\pi/3))
=2^{-4/9}(1-t)^{1/9}[1+2^{-4/3}(1-t)^{1/3}-\frac{5}{72}(1-t)\nonumber\\
&&\hspace{1in}+\frac{2^{1/3}}{128}(1-t)^{5/3}+\frac{7\cdot 2^{2/3}}{288}(1-t)^{4/3}+O(t^2)]
\end{eqnarray}
 
These results for $\lambda=\cos(\pi/4)~{\rm and}~\cos(\pi/3)$ all
  agree with (\ref{res1})-(\ref{res3}).

\subsection{The Ising case $\lambda=1$ ($\sigma=0$)}
The Ising case $\lambda=1$ has been extensivly studied in \cite{ongp}
where it is shown that as $t\rightarrow 1$
\begin{eqnarray}
\hspace{-.8in}C^-(N,t;1) =C(N,t=1)[1
-\frac{N}{4}(1-t)\left(\ln(1-t)-\ln16+\sum_{n=1}^N n^{-1}\right)+\cdots]
\label{isingatt1ngen}
\end{eqnarray}
with
\begin{equation}
\hspace{-.8in}C(N,t=1;1)=\left(\frac{2}{\pi}\right)^N
\prod_{m=1}^{N}\left(1-\frac{1}{4m^2}\right)^{m-N}
\label{ct1sig0alln}
\end{equation}
which agrees with (\ref{res1})-(\ref{res3}) in the limit
$\sigma\rightarrow 0$.

\subsection{The case $\lambda=0$ ($\sigma=1$)}

When $\sigma=1$ we find from (\ref{res2}) and (\ref{res3}) that
\begin{equation}
{\hat s}(N,1)=\frac{16}{2N+1},~~~K(N,1)=\frac{1}{2}
\end{equation} 
and thus (\ref{cmat1final}) reduces to
\begin{equation}
C^-(N,t;0)=(1-t)^{1/4}
\end{equation}
as required by (\ref{fred1}).

\subsection{$N \rightarrow \infty$ for $K(N,\sigma)$}

To obtain the behavior of $K(N;\sigma)$ for $N\rightarrow \infty$ we
use the identity
\begin{equation}
\frac{\sin \pi \delta}{\pi \delta}
=\prod_{m=1}^{\infty}\left(1-\frac{\delta^2}{m^2}\right)
\end{equation}
to write
\begin{eqnarray}
\hspace{-,2in}\frac{\sin \pi \sigma/2}{\sigma}
=\frac{\pi}{2}\prod_{m=1}^{\infty}\left(1-\frac{\sigma^2}{4m^2}\right)
=\prod_{l=1}^{\infty}\left(1-\frac{1}{4l^2}\right)^{-1}
\prod_{m=1}^{\infty}\left(1-\frac{\sigma^2}{4m^2}\right)
\end{eqnarray}
which we use in (\ref{res3}) to obtain
\begin{eqnarray}
&&K(N;\sigma)=2^{-\sigma^2}\prod_{m=1}^{N-1}\left(1-\frac{\sigma^2}{4m^2}\right)^{-m}
\prod_{m=N}^{\infty}\left(1-\frac{\sigma^2}{4m^2}\right)^N\nonumber\\
&&\hspace{.8in}\times\prod_{l=1}^{N-1}\left(1-\frac{1}{4l^2}\right)^{l}
\prod_{l=N}^{\infty}\left(1-\frac{1}{4l^2}\right)^{-N}
\end{eqnarray}
To now expand $K(N;\sigma)$ for $N\rightarrow \infty$ we use for the
products running $N$ to infinity
\begin{eqnarray}
&&ln
\prod_{m=N}^{\infty}\left(1-\frac{\sigma^2}{4m^2}\right)^N
=N\sum_{m=N}^{\infty}\ln (1-\sigma^4/4m^2)\nonumber\\
&&\sim
-N\sigma^2/4\int_N^{\infty}dm/m^2=-\sigma^2/4
\end{eqnarray}
For the  products from $1$ to $N-1$ we write
\begin{equation}
\prod_{m=1}^{N-1}\left(1-\frac{\sigma^2}{4m^2}\right)^{-m}
=\prod_{m=1}^{N-1}e^{\sigma^2/4m}
\prod_{m=1}^{N-1}\left(1-\frac{\sigma^2}{4m^2}\right)^{-m}e^{-\sigma^2/4m}
\end{equation}
where for  $N\rightarrow \infty$ the second product converges and the first
product is expanded using the definition of Eulers constant $\gamma$
\begin{equation}
\lim_{N\rightarrow \infty}\left(\sum_{n=1}^N\frac{1}{n}-\ln N \right)=\gamma
\end{equation} 
to find for $N\rightarrow \infty$
\begin{equation}
\ln \prod_{m=1}^{N-1}e^{\sigma^2/4m}=
\frac{\sigma^2}{4}\sum_{m=1}^{N-1}\frac{1}{m}\rightarrow
\frac{\sigma^2}{4}(\ln N+\gamma)
\end{equation}
Thus we have for $N\rightarrow \infty$
\begin{eqnarray}
&&K(N;\sigma)\rightarrow
N^{(\sigma^2-1)/4}2^{-\sigma^2}e^{-(\sigma^2-1)(1+\gamma)/4}\nonumber\\
&&\hspace{.5in}\times\prod_{m=1}^{\infty}\left(1-\frac{\sigma^2}{4m^2}\right)^{-m}e^{-\sigma^2/4m}
\prod_{m=1}^{\infty}\left(1-\frac{1}{4m^2}\right)^{m}e^{1/4m}
\end{eqnarray}
When this is rewritten in terms of Barnes G functions and the
derivative of the zeta function at $-1$ this agrees with the result
obtained by Tracy \cite{tracy} for the scaling limit of $C^-(N,t;\lambda)$

\section{The Toda-like equation}

In 2010 Mangazeev and Guttmann \cite{mangazeev} proved that
$C^-(N,t;\lambda)$ satisfies the following Toda-like equation
\begin{eqnarray}
\hspace{-.9in}(1-t)^2\frac{d}{dt}t\frac{d}{dt}\ln C^-(N.t;\lambda)+N^2 =
(N^2-1/4)
\frac{C^-(N+1,t;\lambda)C^-(N-1,t;\lambda)}{C^-(N,t;\lambda)^2}
\label{toda}
\end{eqnarray}
The verification of this identity in the limit $t\rightarrow 1$ using
the expansion (\ref{cmat1final}) with (\ref{res1})-(\ref{res3}) combined with
the previous results for $N=0,~1$ constitutes an
inductive proof of the connection formulas (\ref{res2}) 
for ${\hat s}(N,\sigma)$ and (\ref{res3}) for $K(N;\sigma)$.

It is straigntforward to see from (\ref{res3}) that
\begin{equation}
\frac{K(N+1;\sigma)K(N-1;\sigma)}{K(N;\sigma)^2}=\frac{N^2-\sigma^2/4}{N^2-1/4}
\end{equation}
and thus for $x=1-t\rightarrow 0$ the right hand side of (\ref{toda})  to
order $x^{1-\sigma}$ is
\begin{eqnarray}
&&\hspace{-.6in}(N^2-1/4)\frac{C^-(N+1,t;\lambda)C^-(N-1,t;\lambda)}{C^-(N,t;\lambda)^2}\nonumber\\
&&\hspace{-.6in}=(N^2-\sigma^2/4)\{1+\frac{x^{1-\sigma}}{16\sigma}[{\hat s}(N+1;\sigma)(2N+2+\sigma)
+{\hat s}(N-1;\sigma)(2N-2+\sigma)\nonumber\\
&&\hspace{1.5in}-2{\hat s}(N;\sigma)(2N+\sigma)]+O(x)\}
\end{eqnarray} 
 
Using (\ref{cmat1final}) we find to order $O(x^{1-\sigma})$ that
\begin{eqnarray}
&&(1-t)^2\frac{d}{dt}t\frac{d}{dt}\ln C^-(N.t;\lambda)+N^2=\nonumber\\
&&\hspace{.5in}N^2-\sigma^2/4-\frac{{\hat s}(N;\sigma)}{16}
(1-\sigma)(2N+\sigma)x^{1-\sigma}+O(x)
\end{eqnarray}
and thus the leading order terms in (\ref{toda}) cancel because of the
connection formulas (\ref{res3}) and from the terms of order
$x^{1-\sigma}$ we obtain the recursion relation which must be
satisfied by ${\hat s}(N;\sigma)$
\begin{eqnarray}
&&-{\hat s}(N;\sigma)\sigma(1-\sigma)(2N+\sigma)\nonumber\\
&&=(N^2-\sigma^2/4)\{{\hat s}(N+1;\sigma)(2N+2+\sigma)
+{\hat s}(N-1;\sigma)(2N-2+\sigma)\nonumber\\
&&\hspace{1in}-2{\hat s}(N;\sigma)(2N+\sigma)\}
\end{eqnarray}
which by direct substitution  is easily seen to be satisfied by the 
expression (\ref{res2})
for ${\hat s}(N;\sigma)$. Thus we have proven by induction that (\ref{res2})
and (\ref{res3}) are correct.


\section*{References}

\begin{thebibliography}{99}


\bibitem{lm} I. Lyberg and B.M. McCoy, Form factor expansion of the
  row and diagonal correlations functions of the two dimensional ising
  model, J. Phys. A 40 (2007) 3329-3346.


\bibitem{wmtb} T.T. Wu, B.M. McCoy, C.A. Tracy and E. Barouch, The
  spin-spin correlation function of the 2-dimensional Ising model:
  Exact results in the scaling region, Phys. Rev. B13 (1976) 316.

\bibitem{book} B.M. McCoy and T.T. Wu, The Two Dimensional Ising
  Model, (Harvard Univ.Press 1973).

\bibitem{wu} T.T. Wu, Theory of Toeplitz determinants and the spin
  correlations of the two-dimensional Ising model, Phys. Rev. 149 (1966) 380.


\bibitem{jm} M. Jimbo and T. Miwa, Studies on holonomic quantum fields
  XVII, Proc. Japan Acad. Ser A Math. Sci. 56 (1980) 405-410; 57
  (1981) 347,


\bibitem{mangazeev} V.V. Mangazeev and A.J. Guttmann, Form factor
  expansion in the 2D Ising model and Painlev{\'e} VI,
  Nucl. Phys. B838 (2010) 391-412


\bibitem{mtw} B.M. McCoy, C.A. Tracy and T.T. Wu, Painlev{\'e} functions of
    the third kind, J. Math. Phys. 18 (1977) 1058-1092.



\bibitem{tracy}
C.A. Tracy, Asymptotics of a $\tau$-function arising in the 
two-dimensional Ising model, Comm. Math. Phys. 142 (1991) 297-311.



\bibitem{jimbo} M. Jimbo, Monodromy problem and boundary conditions for
  some Painlev{\'e} equations, Publ. Rims, Kyoyo Univ. 18 (1982) 1137-1161.

\bibitem{its} A.R. Its, O. Lisovyy and A. Prokhorov, Monodromy
  dependence and connection formulae for isomonodromic tau functions,
  Duke. Math. J. (2018) DOI 10.1215/00127094-2017-0055.

\bibitem{guzzetti} D. Guzzetti, The logarithmic asymptotics of the
  sixth Painlev{\'e} equation, J. Phys. A 41 (2008) 205201 (46pp).


\bibitem{ongp} W.P  Orrick, B. Nickel, A.J. Guttmann and J.H.H. Perk,
  The susceptibility of the square latice Ising model: New
  developments, J. Stat. Phys. 102 (2001) 795-841.


\bibitem{boukraa} S. Boukraa, S. Hassani, J-M. Maillard, B.M. McCoy,
  W.P. Orrick and N. Zenine, Holonomy of the Ising model form factors,
  J. Phys. A 40 (2007) 75-111.



\bibitem{mccoyetal}
B.M. McCoy, M. Assis, S Boukraa, S. Hassani, J-M Maillard, W.P. Orick
and N. Zenine, The saga of the Ising model, Publ. Rims. Kyoto Univ. 46 (2010), 
arXiv:1003.0751v2

\bibitem{bateman2} E. Erd{\'e}lyi et al, Higher Transcendental Functions,
  Vol.2 (McGraw-Hill 1955).

\end{thebibliography}
\end{document}